\documentclass[prl,aps,twocolumn,superscriptaddress]{revtex4}
\usepackage{amsmath,graphicx,epsfig}

\usepackage{euscript}
\usepackage{amsfonts}
\usepackage{amssymb}
\usepackage{float}

\def\prn#1{{\left(#1\right)}}

\def\ket#1{{|#1\rangle}}
\def\bra#1{{\langle#1|}}

\def\cg(#1,#2)(#3,#4)(#5,#6){\bra{#1,#2,#3,#4}#5,#6\rangle}

\def\ts#1{{_{\mbox{\scriptsize #1}}}}

\def\threej(#1,#2)(#3,#4)(#5,#6){\begin{pmatrix}#1&#3&#5\\#2&#4&#6\end{pmatrix}}
\def\sixj(#1,#2,#3)(#4,#5,#6){\begin{Bmatrix}#1&#2&#3\\#4&#5&#6\end{Bmatrix}}
\def\ninej(#1,#2,#3)(#4,#5,#6)(#7,#8,#9){\begin{Bmatrix}#1&#2&#3\\#4&#5&#6\\#7&#8&#9\end{Bmatrix}}

\def\sV{{\ensuremath{\EuScript V}}}

\newcommand{\Sr}{$^{88}$Sr$^+$~}
\newcommand{\Parity}{{\Pi}}

\newcommand{\Pin}{P_{\textrm{init}}}

\newcommand{\up}{{\ket{\uparrow}}}
\newcommand{\down}{{\ket{\downarrow}}}

\newcommand{\kupdown}{{\ket{\uparrow\downarrow}}}
\newcommand{\kdownup}{{\ket{\downarrow\uparrow}}}

\newcommand{\upup}{{\uparrow\uparrow}}
\newcommand{\updown}{{\uparrow\downarrow}}
\newcommand{\downup}{{\downarrow\uparrow}}
\newcommand{\downdown}{{\downarrow\downarrow}}

\def\mr{\mathrm}
\def\mb{\mathbf}

\newlength{\defbaselineskip}
\newcommand{\setlinespacing}[1]%
           {\setlength{\baselineskip}{#1 \defbaselineskip}}

\begin{document}

\title{Constraints on exotic dipole-dipole couplings between electrons at the micrometer scale} 
\author{Shlomi Kotler}
\affiliation{National Institute of Standards and Technology, 325 Broadway St., Boulder CO, 80305, USA.}
\author{Roee Ozeri}
\affiliation{Department of Physics of Complex Systems, Weizmann Institute of Science, P. O. Box 26, Rehovot 76100, Israel.}
\author{Derek F. Jackson Kimball}
\affiliation{Department of Physics, California State University - East Bay, Hayward, California 94542-3084, USA.}

\date{\today}



\begin{abstract}
New constraints on exotic dipole-dipole interactions between electrons at the micrometer scale are established, based on a recent measurement of the magnetic interaction between two trapped \Sr ions. For light bosons (mass $\le$ 0.1 eV) we obtain a $90\%$ confidence interval on axial-vector mediated interaction strength of $\left|g_A^eg_A^e/4\pi\hbar c\right|\le 1.2\times 10^{-17}$. Assuming CPT invariance, this constraint is compared to that on anomalous electron-positron interactions, derived from positronium hyperfine spectroscopy. We find that the electron-electron constraint is six orders of magnitude more stringent than the electron-positron counterpart. Bounds on pseudo-scalar mediated interaction as well as on torsion gravity are also derived and compared with previous work performed at different length scales. Our constraints benefit from the high controllability of the experimental system which contained only two trapped particles. It therefore suggests a useful new platform for exotic particle searches, complementing other experimental efforts.
\end{abstract}



\maketitle

Extensions of the standard model which predict new particles may involve modifications to the known spin-spin interaction between fermions~\cite{Dob06}. Examples include pseudo-scalar fields (such as the axion~\cite{Moo84}) which naturally emerge from theories with spontaneously broken symmetries~\cite{Wei78,Wil78,Pec77}, and axial-vector fields such as paraphotons~\cite{Dob05} and extra Z bosons~\cite{Dob06,App03}, which appear in new gauge theories. Both pseudo-scalar and axial-vector fields are candidates to explain dark matter~\cite{Gia05}, dark energy~\cite{Fri03,Fla09}, mysteries surrounding CP-violation~\cite{Moo84}, the hierarchy problem~\cite{Gra15}, and are a generic prediction of string theories~\cite{Arv10}.

Typically, to constrain the effect of a new particle of mass $m$ it is favorable to perform an experimental investigation at a length scale $\lambda\equiv\hbar/mc$, the reduced Compton wavelength associated with the particle. Here, $\hbar$ is Planck's constant divided by $2\pi$ and $c$ is the speed of light. At the macroscopic scale, there have been a number of searches for exotic dipole-dipole interactions between electrons \cite{Vor88,Rit90,Bob91,Win91,Chu93,Hun13,Hec13}, ranging from distance scales of a few cm to the radius of the Earth. The large scales involved contributed to the exquisite sensitivity of these endeavors, allowing for signal averaging over a large number of spins. Such an approach, however, cannot be implemented for scales significantly smaller than a millimeter. Attempting to extrapolate these bounds to the micrometer scale also fails, due to the interactions' distance scaling~\cite{footnote:scaling}.

The magnetic dipole-dipole interaction between two electron spins was measured directly for the first time only recently~\cite{Kot13}, at a previously inaccessible length scale of $r=2.4~{\rm \mu m}$. Here we exploit this recent measurement to constrain exotic spin-dependent interactions between electrons, considerably improving on previous bounds at the micrometer scale.


The new measurement reported in Ref.~\cite{Kot13} benefits from the high controllability of trapped ions and the resulting relatively straightforward analysis of systematic errors. Briefly, two $^{88}$Sr$^+$ ions are trapped in a harmonic potential [$\omega_{\mr{trap}}=2\pi\times 2.386(2)\ {\rm MHz}$] using a linear radio-frequency Paul trap. After laser cooling the ions form a Coulomb crystal with a separation of $r=2.407(1)\ \mu\mr{m}$. Each \Sr ion has a single valence electron. A magnetic field of $\sim 0.47~{\rm mT}$ sets the quantization axis along the line connecting the two ions. The two electrons' state is initialized to $\kupdown$ with $\Pin>0.98$ fidelity, where $\uparrow$($\downarrow$) indicates a spin polarized along the positive(negative) magnetic field direction. The spins evolve under the spin-spin interaction for $T=15$ s, ideally resulting in an entangled state $\ket{\Psi(T)}=\cos(2\xi T)\kupdown+i\sin(2\xi T)\kdownup$. Finally the coherence between $\kupdown$ and $\kdownup$ is quantified by rotating the spins collectively $\up\mapsto (\up+\down)/\sqrt{2},\down\mapsto (\up-\down)/\sqrt{2}$ and measuring the parity observable, $\Parity\equiv P_{\upup}+P_{\downdown}-P_{\updown}-P_{\downup}$, where $P_{ij}$ represents the probabilities to measure the spin system in respective states $\ket{ij}$. Since the parity observable is linear in $\sin(4\xi T)$, the coupling strength $\xi$ can be extracted: $\xi=2\pi\times1.020(95)_{\mr{stat}}\ \textrm{mHz}$, where the error is dominated by statistical uncertainty. This approach allows the experimental observable to be insensitive to spatially homogeneous magnetic field noise.  The experiment also employed a spin-echo pulse technique~\cite{Kot11} to reduce the effect of magnetic field gradients to negligible levels. The measurement of $\xi$ thus gives the interaction strength between the two bound valence electrons of two separate \Sr ions. The gyromagnetic ratio $g$ of these bound electrons, in the presence of the field of the nucleus and the core electrons, differs from that of free electrons, $g_{\mr{free}}$, by $(g_{\mr{free}}-g)/g_{free}\sim 1.8\times 10^{-5}$ (Breit formula~\cite{Bre28}). Therefore, the fractional difference between the measured result and the result that would be obtained for two free electrons can be estimated to be $3.55\times 10^{-5}$, which is well below the measurement statistical uncertainty. This is also the case for all other considered systematic errors in the experiment, which are analyzed and detailed in the supplementary information of Ref.~\cite{Kot13}. 

The measured magnetic interaction strength can easily be compared with the calculated magnetic dipole-dipole interaction between two electrons, 
\begin{equation}
\xi_{\textrm{theory}}=\frac{\mu_0}{4\pi \hbar } \frac{\mu_B^2}{r^3}\left(\frac{g}{2}\right)^2,
\end{equation}
where $\mu_0$ is the magnetic permeability of vacuum and $\mu_B$ is the Bohr magneton. This leads to a theoretical estimate of $\xi_{\textrm{theory}}=2\pi\times0.931(1)\ \textrm{mHz}$. The main source of error in determining $\xi_{\rm{theory}}$ is the statistical uncertainty in $r$ which is estimated from the measured trap frequency, $\omega_{\mr{trap}}$. Thermal fluctuations of $r$ average out to leading order, since they occur at the trap frequency, and are therefore negligible. The agreement of the measured value $\xi=2\pi\times 1.020(95)_{\mr{stat}}\ \textrm{mHz}$ with the predicted value $\xi_{\mr{theory}}$ stands at $\Delta\xi/2\pi=208\ \mu{\rm Hz}$ at the 90\% confidence level. 

This small uncertainty in radial frequency, $\Delta\xi$, translates into new constraints for the strength of axial-vector mediated interactions between electrons. According to Dobrescu and Mocioiu \cite{Dob06} the exchange of a new vector/axial-vector $A$ between fermions results in a Yukawa-type potential~\cite{Dob06,Moo84},

\begin{equation}
\sV_A(r)  = \frac{g_A^e g_A^e}{4 \pi \hbar c}  \frac{\hbar c}{r} \mathbf{S}_1 \cdot \mathbf{S}_2 e^{-r / \lambda } ~, \label{Eq:V2}
\end{equation}
where $g_A^e g_A^e/\prn{ 4 \pi \hbar c }$ is the dimensionless axial-vector coupling constant between the electrons, $\mathbf{S}_{1,2}$ are the electron spins and $r$ is the inter-particle separation.


The $\sV_{A}$ potential would contribute to the $\kupdown\leftrightarrow\kdownup$ coherent oscillation measured in Ref.~\cite{Kot13}, according to the corresponding Hamiltonian, written in the $\kupdown,\kdownup$ basis, 
\begin{equation}
H_{A}(r) = \frac{g_A^e g_A^e}{4 \pi \hbar c}  \frac{2\hbar c}{r} e^{-r / \lambda } \begin{pmatrix}0&1\cr 1&0\cr\end{pmatrix}.
\end{equation}

\begin{figure}[!hbtp]
	\center
	\includegraphics[width=3.35 in]{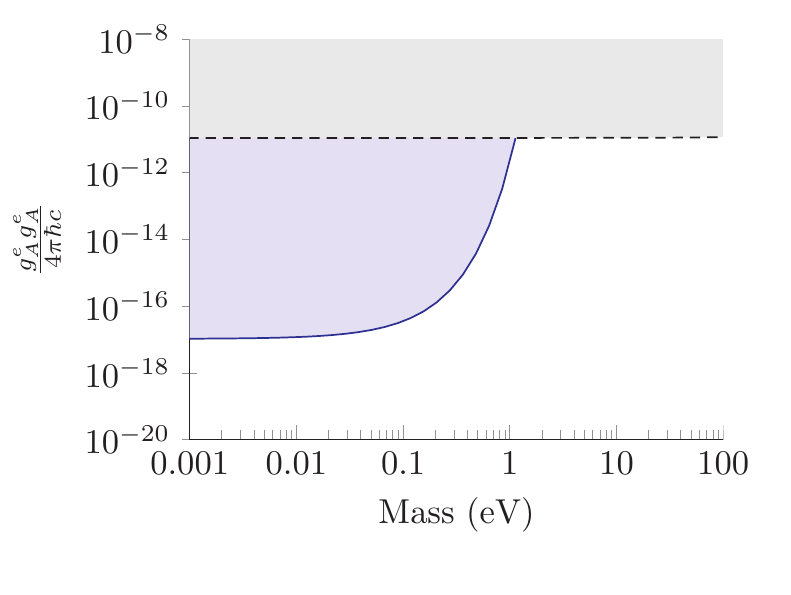}
	\caption{Constraints (at the 90\% confidence level) on the dimensionless coupling constants, $g_A^e g_A^e/\prn{ 4 \pi \hbar c }$ and $g_A^e \bar{g}_A^e/\prn{ 4 \pi \hbar c }$ as a function of the corresponding boson mass. Filled areas correspond to excluded values.  The solid line and dark blue fill show the constraints derived from the possible contribution of an axial-vector mediated interaction to the dipole-dipole interaction between electrons measured in Ref.~\cite{Kot13}.  The dashed line and light gray region show the constraints on the strength of axial-vector mediated electron-positron interaction, $g_A^e \bar{g}_A^e/\prn{ 4 \pi \hbar c }$, by comparing the measurement of the ground-state hyperfine interval for positronium~\cite{Ish14} to recent QED calculations \cite{Eid14}.\label{Fig:ee-axial-constraints}}
\end{figure}

The corresponding oscillation frequency, 
\begin{equation}\xi_{A} =\frac{g_A^e g_A^e}{4 \pi \hbar c}  \frac{c}{r} e^{-r / \lambda }\end{equation}
is smaller than $\Delta\xi=2\pi\times 208\ {\rm \mu Hz}$ for $r=2.4~{\rm \mu m}$, at the 90\% confidence level. This leads to the constraints plotted in Fig.~\ref{Fig:ee-axial-constraints} (dark blue region). 


As seen from Eq. \ref{Eq:V2}, the ability to constrain $\sV_{A}$ greatly benefits from the small experimental scale $r$ at which the experiment is performed. Previous measurements performed at scales much greater than a micrometer render much weaker bounds. Therefore, our bounds (dark blue region, Fig.~\ref{Fig:ee-axial-constraints}) need only be compared to those derivable from smaller scale measurements. Specifically, we compare our result to measurement of the hyperfine structure interval of the ground state in positronium (Ps) \cite{Les14}, which constrains exotic dipole-dipole interactions between electrons and positrons above the angstrom scale (light gray region, Fig. \ref{Fig:ee-axial-constraints}). 

The strength of the axial-vector interaction at the angstrom scale is four orders of magnitude larger than at the micron scale due to the $1/r$ dependence of the interaction (Eq.~\ref{Eq:V2}). Our \Sr based constraints still improve on the Ps measurements due to the exquisite sensitivity of the \Sr experiment, measuring energy shifts with an uncertainty some ten orders of magnitude smaller than that obtained with Ps spectroscopy. To see this, we note that the present agreement between experimental measurements~\cite{Mil75,Rit84,Ish14}, $\Delta E_{HF}({\rm Ps})\ts{expt} = 203~394.2 (1.6)\ts{stat} (1.3)\ts{sys}~{\rm MHz}$, and the most recent theoretical calculations based on quantum electrodynamics (QED)~\cite{Eid14}, $\Delta E_{HF}({\rm Ps})\ts{theory} = 203~391.90 (25)~{\rm MHz}$, is $5~{\rm MHz}$ at the 90\% confidence level~\cite{footnote2}, dominated by the experimental uncertainty.

Similar to Ref.~\cite{Kar10,Les14}, we can estimate the shift of positronium's hyperfine structure interval due to $\sV_A(r)$ using first-order perturbation theory. In this case, one must average over the spherically symmetric ground state (see, for example, Refs.~\cite{Kim10,Led13}) $\ket{\psi}=e^{-r/2a_0}/\sqrt{8\pi a_0^3}$, where $a_0$ is the Bohr radius. 
We obtain for the associated energy shift of the Ps ground-state hyperfine structure interval,
\begin{align}
\Delta E_A = \frac{g_A^e \bar{g}_A^e}{4 \pi \hbar c} \frac{\hbar c}{2a_0} \frac{1}{\prn{ 1+a_0/\lambda }^2}~,
\end{align}
As shown in Ref.~\cite{Kar10}, demanding that $\left|\Delta E_A\right| < 2\pi\hbar\times 5~{\rm MHz}$ leads to the constraints shown in Fig.~\ref{Fig:ee-axial-constraints} (light gray region). These turn out to be essentially mass independent, since in our region of interest, $\lambda\gg a_0$.

A similar comparison can be made for a pseudo-scalar mediated interaction. A new axion-like pseudo-scalar ($P$) particle results in a dipole-dipole potential between electrons \cite{Dob06,Moo84},

\begin{multline}\label{Eq:V3}
\sV_P(\mathbf{r}) = \frac{g_P^e g_P^e}{4 \pi \hbar c} \frac{\hbar^3}{4 m_e^2 c} \left[ \mathbf{S}_1 \cdot \mathbf{S}_2 \prn{ \frac{1}{\lambda r^2} + \frac{1}{r^3} + \frac{4\pi}{3} \delta^3(\mb{r})} \right. \\
\left. - \prn{ \mathbf{S}_1 \cdot \hat{ \mathbf{r} } } \prn{ \mathbf{S}_2 \cdot \hat{ \mathbf{r} } }  \prn{  \frac{1}{\lambda^2 r} + \frac{3}{\lambda r^2} + \frac{3}{r^3} } \right] e^{ - r / \lambda }~,
\end{multline}
where $g_P^e g_P^e/\prn{ 4 \pi \hbar c }$ is the dimensionless pseudo-scalar coupling constant between the electrons, $m_e$ the electron mass and $\hat{r}$ is the unit vector along the line connecting the two fermions~\cite{footnote1}.

\begin{figure}[!hbtp]
	\center
	\includegraphics[width=3.35 in]{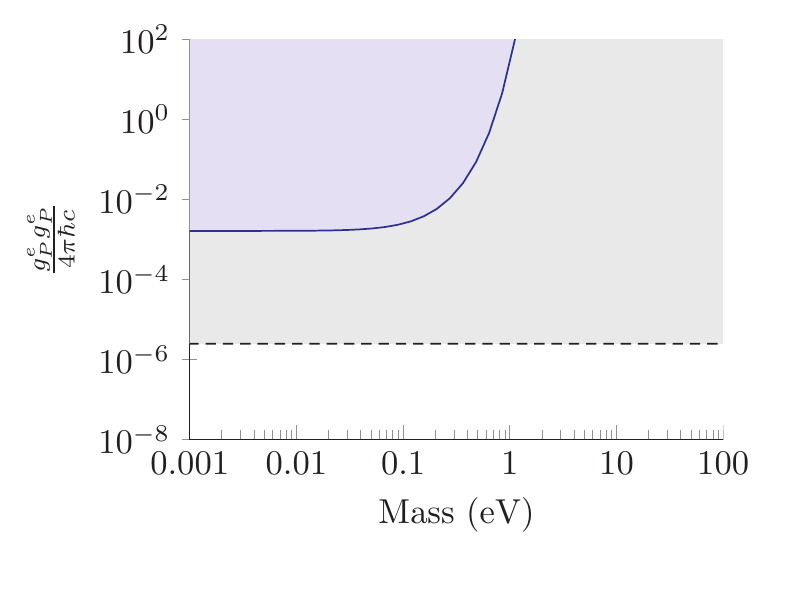}
	\caption{Same as Fig.~\ref{Fig:ee-axial-constraints}, for the case of a pseudo-scalar mediated interaction, constraining $g_P^e g_P^e/\prn{ 4 \pi \hbar c }$ for electron-electron (dark blue region) and $g_P^e \bar{g}_P^e/\prn{ 4 \pi \hbar c }$ for electron-positron (light gray region).\label{Fig:ee-pseudo-constraints}}
\end{figure}

Here, the \Sr experimental sensitivity is not sufficient to overcome the larger pseudo-scalar interaction strength at the angstrom scale probed by the Ps measurement, due to the $1/r^3$ dependence of the potential. For the \Sr experiment, $\sV_P$ would contribute to the $\kupdown\leftrightarrow\kdownup$ coherent oscillation measured in Ref.~\cite{Kot13} by
\begin{equation}\xi_{P} =\frac{g_P^e g_P^e}{4 \pi \hbar c} \frac{\hbar^2}{4 m_e^2 c} \prn{\frac{1}{r^3}+\frac{1}{\lambda r^2}}e^{-r/\lambda}.
\end{equation}
Demanding that $\xi_P$ be smaller than $\Delta\xi=2\pi\times 208\ {\rm \mu Hz}$ for $r=2.4~{\rm \mu m}$ at the 90\% confidence level leads to the constraints plotted in Fig.~\ref{Fig:ee-pseudo-constraints} (dark-blue region). In comparison, the energy shift of the Ps ground-state hyperfine structure interval due to $\sV_P(r)$ is
\begin{align}
\Delta E_P = \frac{g_P^e \bar{g}_P^e}{4 \pi \hbar c} \frac{\hbar^3}{24 m_e^2c a_0^3}  \prn{ 1 - \frac{1}{\prn{ 1 + \lambda/a_0 }^2}}~.
\end{align}
As shown in Ref.~\cite{Les14} demanding that $\left|\Delta E_P\right| < 2\pi\hbar\times 5~{\rm MHz}$ leads to the constraints depicted by the light gray region in Fig.~\ref{Fig:ee-pseudo-constraints}, again, as in Fig.~\ref{Fig:ee-axial-constraints}, showing nearly mass-independent behavior over the depicted range.

We note that direct comparison between the electron-electron and electron-positron constraints is based on the implicit assumption of CPT-invariance for the exotic interactions studied here. The potentials $\sV_A(r)$ and $\sV_P(r)$ are even under the parity (P) and time-reversal (T) transformations, and thus they are also even under the combined PT symmetry. Consequently, assuming CPT-invariance (where C represents charge conjugation), electrons and positrons should have the same magnitude of coupling strength to exotic pseudoscalar or axial-vector interactions: $|g_P^e| = |\bar{g}_P^e|$ and $|g_A^e| = |\bar{g}_A^e|$. It should be noted, however, that CPT invariance is not guaranteed for these exotic interactions, at least based on the current state of knowledge~\cite{Alt13}. Thus, one can regard the $e^--e^-$ and the $e^+-e^-$ measurements discussed here as independent constraints. 

The above analysis can also be used to place bounds on torsion gravity.  According to general relativity, the local space-time curvature is unaffected by the presence of spin \cite{Kob62,Lei64,Heh90,Khr98,Sil05}.  However, in extensions of general relativity based on a Riemann-Cartan spacetime, the gravitational interaction is described by a torsion tensor which can generate spin-mass and spin-spin interactions \cite{Heh76,Sha02,Ham02,Kos08,Obu14}. The spin-spin interaction generated by a standard propagating torsion field takes the form of $\sV_P(r)$ \cite{Ham95,Ade09}, with $\lambda\to\infty$ and a coupling strength that can be parameterized in terms of a dimensionless parameter $\beta$, such that 
\begin{align}
\beta^2 = \prn{ \frac{g_P^e g_P^e}{4 \pi \hbar c} } \times \prn{ \frac{2}{9} \frac{\hbar c}{G m_e^2} }
\end{align}
where $G$ is the gravitational constant. The minimally coupled Dirac equation for a spin-$1/2$ particle \cite{Nev80,Nev82,Car94,Ham95} predicts $\beta = 1$. Experimental constraints, based on the analysis presented here, are shown in table~\ref{tbl:torsion}. 

\begin{table}[!h]
	\begin{ruledtabular}
    \begin{tabular}{llll}
	Particles	& Experiment 	& $\left|\beta^2\right|\le$	& Ref.\\
				& Scale $\ge$	&				&			\\
	\hline
	$n-n$    	& 50 cm 	& $2\times 10^{28}$	& \cite{Vas09}\\ 	
	$p-p$		& 50 cm		& $2\times 10^{31}$ & \cite{Kim14} (exp.~\cite{Vas09})\\
	$e^--e^-$ 	& 15 cm    	& $4\times 10^{28}$	& \cite{Hec13}\\ 
	$e^--e^-$ 	& 2 $\mu$m 	& $2\times 10^{41}$	& this work (exp.~\cite{Kot13})\\ 
	$n-n$		& 100 nm	& $6\times 10^{34}$	& \cite{Sno11}\\
	$p-p$		& \AA		& $1\times 10^{33}$	& \cite{Ram79}\\
	$e^--e^+$ 	& \AA		& $3\times 10^{38}$	& this work (exp.~\cite{Ish14})\\ 
	$p-p$, $n-p$& \AA 		& $1\times 10^{31}$	& \cite{Led13}\\
	\end{tabular}
	\end{ruledtabular}
	\caption{Comparison of torsion gravity constraints.}\label{tbl:torsion}
\end{table}

Finally we note that stellar energy-loss arguments strongly constrain the pseudoscalar coupling of electrons \cite{Raf12}, in particular revealing that for particles with mass $m \lesssim 10~{\rm keV}$ \cite{Raf95}, $\left|g_P^e g_P^e/(4 \pi \hbar c)\right| \lesssim 10^{-25}$,
far exceeding the laboratory limits discussed here. In terms of torsion gravity, these astrophysical constraints translate to a limit $\beta^2 \lesssim 10^{19}$.
These constraints, however, do not apply to the axial-vector interactions \cite{Dob06}.

In conclusion, the results of a new direct measurement of the magnetic interaction between two electrons at the micrometer scale were used to place bounds on exotic forces. The constraints on an axial-vector mediated spin-spin interaction between electrons are six orders of magnitude more stringent than those derived from positronium spectroscopy for electron-positron pairs, at the micrometer scale (equivalently for masses below $\sim 1\ \mr{eV}$). Note that the constraint could be improved by one or two orders of magnitude, based on an extension of the technology used in Ref.~\cite{Kot13}. This would require a tighter ion trap, consequently placing the two electronic spins at separations less than a micrometer.

\begin{acknowledgments}
The collaboration resulting in this paper was suggested by Dmitry Budker. The authors are sincerely grateful to Mitch Watnik, Joshua Kerr, Gilad Perez, Moty Milgrom and Guy Gur-Ari Krakover, for enlightening discussions. We thank Dmitry Budker and Jennie Guzman for careful reading of our manuscript. This work was supported in part, by the U.S. National Science Foundation under grant PHY-1307507 and by I-Core: the Israeli excellence center ``circle of light" and the European Research Council (consolidator Grant Ionology-616919).
\end{acknowledgments}


\begin{thebibliography}{99}
\bibitem{footnote:scaling} As can be seen from Eq.~(\ref{Eq:V2}) and Eq.~(\ref{Eq:V3}), extrapolating these bounds to the micrometer scale deteriorates their sensitivity roughly by a factor of $e^x/x$, where $x$ is the ratio between the measurement scale and the extrapolated scale, i.e. $x\ge 1\ \mr{ mm}/1\ \mu\mr{m}=1000$. That renders bounds obtained at the millimeter scale useless at the micrometer scale.
\bibitem{footnote2}
Statistical and systematic uncertainties have been combined in quadrature and a normal distribution has been assumed.
\bibitem{footnote1}
Note that Ref.~\cite{Dob06} neglects the term containing the delta-function, since Dobrescu and Mocioiu focus on long-range potentials; the delta-function term is present in the analogous equation from Ref.~\cite{Moo84}.
\end{thebibliography}
\end{document}